\documentclass[nofootinbib,superscriptaddress,twocolumn]{revtex4}
\usepackage[cp1251]{inputenc}
\usepackage[english]{babel}
\usepackage{csquotes}
\usepackage{graphicx}
\usepackage{float}
\usepackage{citehack}
\usepackage{amsmath}
\usepackage{fancyref}
\usepackage{wasysym}
\usepackage{pdfpages}
\usepackage{lineno}

\begin{document}

\title{A search for low-energy neutrinos correlated with gravitational wave events GW150914, GW151226 and GW170104 with the Borexino detector}
\date{\today}
\author{M.~Agostini}
\affiliation{ Gran Sasso Science Institute (INFN), 67100 L'Aquila, Italy}
\author{K.~Altenm\"{u}ller}
\affiliation{Physik-Department and Excellence Cluster Universe, Technische Universit\"at  M\"unchen, 85748 Garching, Germany}
\author{S.~Appel}
\affiliation{Physik-Department and Excellence Cluster Universe, Technische Universit\"at  M\"unchen, 85748 Garching, Germany}
\author{V.~Atroshchenko}
\affiliation{National Research Centre Kurchatov Institute, 123182 Moscow, Russia}
\author{Z.~Bagdasarian}
\affiliation{IKP-2 Forschungzentrum J\"ulich, 52428 J\"ulich, Germany}
%\affiliation{RWTH Aachen University, 52062 Aachen, Germany}
\author{D.~Basilico}
\affiliation{Dipartimento di Fisica, Universit\`a degli Studi e INFN, 20133 Milano, Italy}
\author{G.~Bellini}
\affiliation{Dipartimento di Fisica, Universit\`a degli Studi e INFN, 20133 Milano, Italy}
\author{J.~Benziger}
\affiliation{Chemical Engineering Department, Princeton University, Princeton, NJ 08544, USA}
\author{D.~Bick}
\affiliation{Institut f\"ur Experimentalphysik, Universit\"at Hamburg, 22761 Hamburg, Germany}
\author{G.~Bonfini}
\affiliation{INFN Laboratori Nazionali del Gran Sasso, 67010 Assergi (AQ), Italy}
% ...

\collaboration{\bf{The Borexino collaboration}}
\begin{abstract}
	We present the results of a low-energy neutrino search using the Borexino detector in coincidence with the gravitational wave (GW) events GW150914, GW151226 and GW170104.
	We searched for correlated neutrino events with energies greater than 250 keV within a time window of $\pm500$~s centered around the GW
	detection time. A total of five candidates were found for all three GW150914, GW151226 and GW170104.
	This is consistent with the number of expected solar neutrino and background events.
	As a result, we have obtained the best current upper limits on the GW event neutrino fluence of all flavors ($\nu_e, \nu_{\mu}, \nu_{\tau}$) in the energy range $(0.5 - 5.0)$~MeV.
\end{abstract}
\maketitle
\section{Introduction}
The observation of two gravitational wave events GW150914 and GW151226 and the candidate LVT151012 by the LIGO experiment \cite{GW150914, GW151226, LIGO} triggered an intensive follow-up campaign in neutrino detectors \cite{IceCube,Auger, KL,SK}.
\v Cerenkov neutrino telecopes (ANTARES, IceCube \cite{IceCube}) and Pierre Auger Observatory \cite{Auger} have searched for high energy neutrinos above $100$~GeV and $100$~PeV respectively.
KamLAND has searched for inverse beta decay (IBD) antineutrino events with energies in the range $(1.8 - 111)$~MeV \cite{KL} and Super-Kamiokande has reported the results for neutrino signals in neutrino energy range from $3.5$~MeV to $100$~PeV \cite{SK}.
The neutrino and antineutrino events within a time window of $\pm 500$ seconds around the gravitational wave detection time were analyzed in the detectors mentioned above, but no evidence for an excess of coincident neutrino events was reported.

Electromagnetic detectors of photons, including X- and $\gamma$-rays \cite{AGILE, TOROS, XMM-Newton, Fermi-LAT, Fermi,Fermi-LAT2} also did not show any counterpart for various wavelengths of electromagnetic radiation except a weak coincident excess above $50$~keV and $0.4$~s after GW159014 claimed by Fermi gamma-ray burst monitor \cite{Fermi, GBM}.
Data from the detectors listed above would be very important in determining the location of the GW source in the sky.

Combination of data from gravitational, neutrino and electromagnetic detectors forms a new multi-messenger approach leading to a more complete understanding of astrophysical and cosmological processes through combination of information from different probes.

Recently LIGO and Virgo Collaboration reported the observation of GW170104, a gravitational-wave signal measured on January 4, 2017 and produced by 50-solar mass binary black hole coalescence \cite{GW170104}. Here we report the results of a search for signals with visible energy above $0.25$~MeV in the Borexino detector in coincidence with GW150914, GW151226 and GW170104 events.
We look for neutrino signals from $\nu_e, \nu_{x=\mu, \tau}$ and antineutrinos $\bar{\nu}_e, \bar{\nu}_{x=\mu,\tau}$ originated in GW events that scatter on electrons.
We also search for signals of $\bar{\nu}_e$ that induce IBD on protons.
Using the unique features of the Borexino detector -- outstanding low background level, large scintillator mass and low energy threshold -- new limits on low-energy neutrino fluence correlated with detected GW events have been obtained.
\begin{section}{Borexino detector}
	
	Borexino is a liquid scintillator detector located underground at $3400$ meters of water equivalent in the Gran Sasso Laboratory, Italy.
	
	\begin{figure}[ht]
		\includegraphics[bb=50 0 410 450,width=0.8\linewidth]{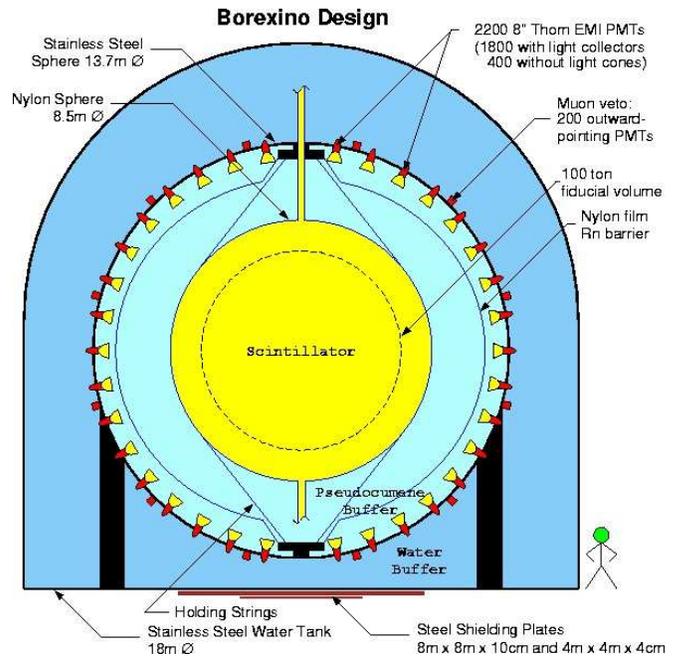}
		\caption{Principal scheme of the Borexino neutrino detector. Doping of PPO and DMP is shown by yellow and cyan colors respectively. Fiducial volume is shown in an arbitrary way and does not reflect the one used in the current analysis.}
		\label{fig:bx_det_constr}
	\end{figure}

	The detector design is based on a concept of graded shielding such that the radio purity level increases moving
	towards the detector center.
	The main housing of the detector is a cylinder with a hemispheric top with a diameter of $18$~m and height of $15.7$~m and is made of stainless steel with high radiopurity.
	Contained inside is a stainless steel sphere (SSS) with a diameter of $6.75$~m and thickness of $8$~mm fixed in place by a stainless steel support structure.
	The space between the outer barrel and stainless steel sphere is filled with ultrapure water and is equipped with $208$ 8-inch PMTs.
	It serves as a \v Cerenkov muon veto and is called the outer detector (OD).
	The inner side of the stainless steel sphere is equipped with $2212$ 8-inch PMTs of the inner detector (ID) and the inner volume is filled with pseudocumene (C$_9$~H$_{12}$).
	The inner detector contains two transparent spherical nylon vessels with a refractive index similar to that of pseudocumene with radii of $5.5$~m (radon barrier) and $4.25$~m (inner vessel, IV) located concentrically within the stainless steel sphere (see Fig.~\ref{fig:bx_det_constr}).
	The nylon used for these vessels was produced underground to fulfill high radiopurity requirements.
	
	The scintillator volume inside the inner vessel has an admixture of PPO used for creation of Stokes shift \cite{Stokes1852}.
	The scintillator outside the inner vessel is doped with DMP that quenches light production decreasing scintillation signals whose origin is not in the IV.
	
	The detector was carefully purified with various liquid handling procedures including water extraction campaign and shows exceptionally low level of radioactive impurities in the bulk of the inner vessel fluid \cite{bx5}.
	
	A detailed description of the detector could be found elsewhere~\cite{bx1, bx2, bx3, bx4, bx5, bx6, bx7, bx8, bx9, bx10, bx11, bx12,bxmc}.
	
	%The uniquely low background of Borexino detector allowed us to measure fluxes of pp-chain neutrinos (\cite{bx3,bx4,bx10,bx13,bx7,bx11,bx12}) and to set the % %tightest upper limit on the flux of solar neutrinos produced in the CNO cycle (\cite{bx11}).
	%The Borexino has set new limits on various rare processes (\cite{bx4,bx15,bx14,bx19,bx17,bx18,antinu_PLB}).

	Borexino first detected and then precisely measured the flux of the $^7$Be solar neutrinos \cite{bx3,bx4,bx13}, has ruled out any significant day-night asymmetry of their interaction rate \cite{bx10}, has measured the $^8$B-neutrino rate with $3$~MeV threshold \cite{bx7}, has made the first direct observation of pep neutrinos \cite{bx11}, has made the first spectral measurement of pp-neutrinos \cite{bx12} and has set the best upper limit on the flux of solar neutrinos produced in the CNO cycle \cite{bx11}.
	The uniquely low background level of the Borexino detector made it possible to set new limits on the effective magnetic moment of the neutrino \cite{bx4}, on the stability of the electron for decay into a neutrino and a photon \cite{bx15}, on the heavy sterile neutrino mixing in $^8$B decay \cite{bx14}, on the possible violation of the Pauli exclusion principle \cite{bx19}, on the flux of high energy solar axions \cite{bx17}, on antineutrinos from the Sun and other unknown sources \cite{antinu_PLB}, on Gamma-Ray bursts neutrino and antineutrino fluences \cite{bx18} and on some other rare processes.
\section{Data Selection}
\begin{figure}[ht]
	\includegraphics[bb=80 75 497 742,width=0.8\linewidth]{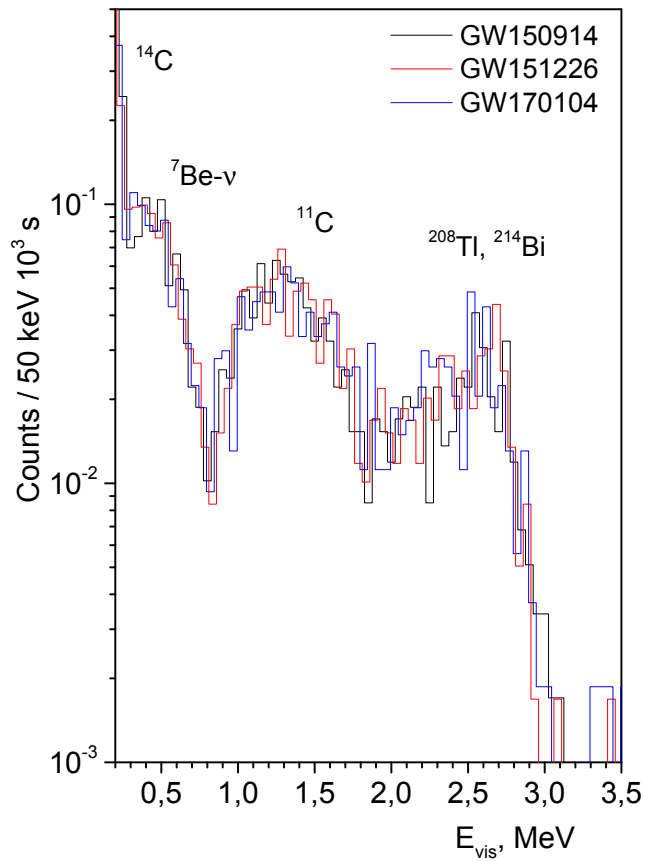}
	\caption{Visible energy spectrum of Borexino detector data passing the selection procedure, obtained using weekly runs containing events GW150914, GW151226 and GW170104. The spectra corresponding to the GW151226 and GW170104 are shifted to the left by 15 and 30 keV for illustrative purposes. The plot shows the main spectral components, such as recoil electrons produced in elastic scattering of solar neutrinos from $^7$Be, decays of cosmogenic $^{11}$C and external gamma events caused by decays of $^{214}$Bi and $^{208}$Tl outside fiducial volume.}
	\label{fig:bx_spectrum}
\end{figure}
The aim of data selection is to provide maximum exposure with minimum  background contribution.
Since the electron neutrino scattering searched in the current analysis has no interaction signature, background reduction has to be performed in a generic manner as a reduction of the detector count rate per unit of exposure.
Thus, one should take the following background sources into consideration:
\begin{itemize}
	\item{Short-lived cosmogenic backgrounds ($\tau \lesssim 1$~s) produced within the detector fiducial volume, such as $^{12}\rm{B}$, $^{8}$He, $^{9}$C, $^{9}$Li etc.}
	\item{Other cosmogenic backgrounds, produced within the detector fiducial volume, such as $^{11}$Be, $^{10}$C, $^{11}$C etc.}
	\item{Backgrounds of the inner nylon vessel, such as $^{210}$Pb and Uranium/Thorium decay chains.}
	\item{Natural backgrounds contained in the bulk of the detector fluid such as $^{14}$C, $^{85}$Kr, $^{210}$Bi and $^{210}$Pb.}
\end{itemize}  

These backgrounds can be suppressed by using information coming from the processed detector data such as ID/OD coincidences and position reconstruction.
Cosmogenic backgrounds can be reduced by applying the detector temporal veto after each muon that could be discriminated through coincidence with outer veto as well as by pulse-shape discrimination \cite{bx9}.
A veto length of $0.3$~s after muons is applied to suppress $^{12}$B to a statistically non-significant level and reduce $^{8}$He, $^{9}$C and $^{9}$Li by factor of 3 with a live time loss of $1$~\%.

Backgrounds contained in the bulk can not be avoided since they can not be localized neither spatially nor temporally.
Nevertheless, the number of counts can be reduced by setting a cut on visible energy.
This is important specifically due to the presence of $^{14}$C in the scintillator.
$^{14}$C produces a beta-spectrum with an endpoint of $0.156$~MeV and has activity of roughly $110$~Bq in the whole inner vessel.

The presence of this spectral component sets the lower threshold of the analysis to $0.25$~MeV of visible energy\footnote{Visible energy spectrum of $^{14}$C  is broadened up to this value due to the detector energy resolution}.
An additional threshold of $0.4$~MeV of visible energy is also used to reduce the $\rm{^{210}Po}$ and $\rm{^{210}Bi}$ background decays and the $^7$Be  solar neutrino scattering on electrons.

Backgrounds contained in the nylon of IV can not be removed by any kind of purification and are therefore of the order of $10^2$ -- $10^3$ times higher than within the bulk of the scintillator.
The most dangerous components are $^{214}$Bi and $^{208}$Tl decays.
These nuclides undergo $\beta$ and $\beta+\gamma$ decay processes with a continuous spectrum overlapping with the region used by this analysis.
The only way to overcome this kind of background is to perform a geometrical cut on events, selecting those within a fiducial volume. The fiducial volume is defined such that all events within and further than $75$~cm away from the IV are kept which corresponds to 3 standard deviations of position reconstruction uncertainty at the lowest energy threshold.\footnote{Position reconstruction precision increases with energy due to statistical reasons}. The corresponding fiducial volume has a mass of 145 t.

The energy spectra after applying these data selection cuts for both weeks containing GW events are shown in Fig.~\ref{fig:bx_spectrum}.
The spectrum is dominated by $^{14}$C in the region below $0.250$~MeV of visible energy, electron recoil from solar $^7$Be neutrinos in $0.25 - 1$~MeV, by cosmogenic $^{11}$C in $1 - 2$~MeV region and by external gamma-quanta of $^{214}$Bi and $^{208}$Tl in $2 - 3$~MeV region.
All these components can not be significantly reduced by any available data selection techniques without serious exposure loss.
The final rates of background events are shown in Table~\ref{Table:GW&Bx Data}.
\section{Analysis and Results}

The observations of GW150914, GW151226 and GW170104 events were made on 14 September and 26 December 2015, and 4 January 2017 respectively, at times when the Borexino detector was taking data.
The detection time and visible energy of Borexino events passing all data selection cuts in $\pm3000$~s windows around GW150914, GW151226 and GW170104 are shown in Fig.~\ref{fig:Fig2}.

A time window of $\pm 500$~s around the GW150914, GW151226 and GW170104 detection times is applied for further analysis.
This time window covers the possible delay of a neutrino which propagates slower than GW (for a claimed distance of $\mathrm{d} \approx 440$ Mpc for  GW151226 \cite{GW151226} the delay reaches $440$~s for a $0.5$~MeV neutrino with 70~meV mass~\footnote{The $Planck$~$2015$ CMB temperature and polarization  power spectra in combination with the baryon acoustic oscillations data gives a limit on the sum of neutrino masses $\sum m_{\nu} \leq 0.17$~eV  at $95$\% C.L. \cite{Plank2015}.  Together with the measured oscillation mass differences \cite{PDG2016} it leads to a constraint on the maximum neutrino mass $m_1$, $m_2$, $m_3$ of $70$~meV.}
), as well as possible earlier emission of neutrinos due to poorly constrained details of black hole - black hole (BH-BH) merger.
Moreover, the choice is consistent with the time window chosen in \cite{IceCube,Auger, KL, SK}.

Two visible energy ranges are used in this analysis, the first is from $0.25$~MeV to $15$~MeV and the second extends from $0.4$ to $15$~MeV.
The lower threshold of $0.25$~MeV allows us to register neutrinos with energy as low as $0.41$~MeV via neutrino-electron elastic scattering.

\begin{figure}[ht!]
	\begin{center}
		\includegraphics[bb=80 50 480 740,width=0.8\linewidth]{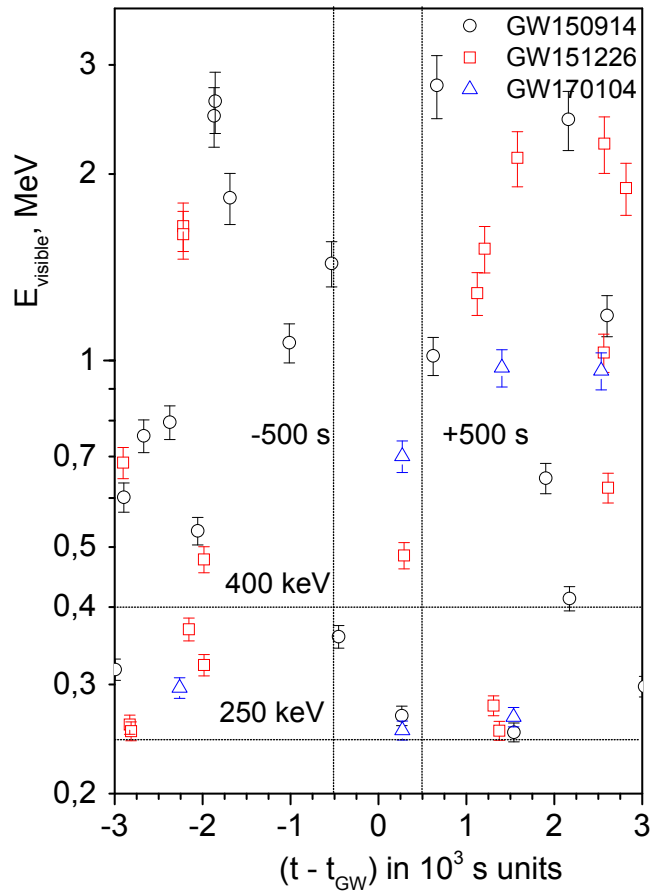}
		\caption{Borexino events between $0.25$~MeV and $15$~MeV visible energy occurring within $\pm 3000$~s of the GW150914 (black circles), the GW151226 (red squares) and the GW170104 (blue triangles) detection times. The closest events with energy $0.267$~MeV, $0.485$~MeV and $0.700$~MeV occurred at $265$~s, $291$~s and $270$~s after the GW150914, the GW151226 and GW17010, correspondingly. All events are consistent with the expected solar neutrino and background count rate.  }
		\label{fig:Fig2}
	\end{center}
\end{figure}

Applying the selection cuts listed above leaves five candidates within $\pm 500$~s search window around the GW150914,  GW151226 and GW170104 detection time respectively(Fig.~\ref{fig:Fig2}). The closest events with energy $0.267$~MeV, $0.485$~MeV and $0.700$~MeV occurred at $265$~s, $291$~s and $270$~s after the GW150914, GW151226 and GW170104, respectively. One should note there are no extra events below 1~MeV within an extended window of $\pm$1000~s. A delay of $1000$~s corresponds to a $70$~meV neutrino which has traveled $990$~Mpc (the distance from GW170104 is $880$~Mpc) with the minimal detectable energy of $0.41$~MeV. 
%Thus, although the delay for a less energetic neutrino could exceed the selected $\pm 500~s$ window boundaries, the conclusions would remain unmodified.  

According to Borexino data from weekly runs containing the GW events, the total number of neutrino and background events expected in three 1000-second time windows is (6.5$\pm$0.1) $(\rm{10^3\: s \: 145\: t)}^{-1}$ and (5.1$\pm$0.1) $\rm{(10^3\: s\: 145\: t)}^{-1}$ for energy intervals $(0.25-15)$~MeV and $(0.4-15)$~MeV, respectively (Table \ref{Table:GW&Bx Data}). 

\begin{table}[!htbp] 
	\begin{center}
		\begin{tabular}{|l|c|c|c|c|c|}
			\hline
			GW event & Threshold,  & Count rate,                    & Detected        \\ 
			&    MeV	   &	ev/1000~s                   &                \\ \hline
			GW150914 &0.25/0.4     & 2.07$\pm$0.06/1.68$\pm$0.06   & 2/0         \\ \hline
			GW151226 &0.25/0.4     & 2.15$\pm$0.06/1.72$\pm$0.06   & 1/1         \\ \hline
			GW170104 &0.25/0.4     & 2.28$\pm$0.07/1.72$\pm$0.06   & 2/1         \\ \hline
			%	Total    &0.25/0.4      & 4.22$\pm$0.08/3.40$\pm$0.08   & 3/1        \\ \hline
		\end{tabular}
		\vskip 12pt
		\caption{Average Borexino count rate in 7 day runs containing the GW events in terms of events per $1000$~s interval for $0.25$~MeV and $0.4$~MeV thresholds. The number of registered events inside $\pm 500$~s interval is shown in the rightmost column.}
		\label{Table:GW&Bx Data}
	\end{center}
\end{table}

The upper limits on the fluence without oscillation for monoenergetic (anti-)neutrinos with energy $E_\nu$ are calculated as follows:
\begin{equation}
\Phi = \frac{N_{90}(E_\nu,n_{obs},n_{bkg})}{\epsilon {N_e\sigma(E_{th},E_\nu)}}
\label{lim}
\end{equation}
where ${N_{90}(E_\nu,n_{obs},n_{bkg})}$ is the $90$~\% C.L. upper limit on the number of GW-correlated events in $(E_{th},E_\nu)$ range per single GW event, $\epsilon$ is the recoil electron detection efficiency, $N_e$ = $4.79\times 10^{31}$ is the number of electrons in the Borexino fiducial volume, $\sigma(E_{th},E_\nu)$ is the total neutrino-electron cross-section integrated over the $(E_{th},E_\nu)$ interval. The recoil electron detection efficiency equals 1 with precision of fiducial volume definition of 4\%.
The upper limit  ${N_{90}(E_\nu,n_{obs},n_{bkg})}$  is calculated for the total number of observed events $n_{obs}$ and for the known mean background $n_{bkg}$ in accordance with the procedure \cite{Fel89}.
The total cross-section $\sigma(E_{th},E_\nu)$ is obtained by integrating the $(\nu,e)$-scattering cross-section $d\sigma(E_\nu)/dE_e$ \cite{bahcall} over recoil electron energies $E_e$ between the electron threshold energy $E_{th}$ and the neutrino energy $E_\nu$:

\begin{equation}
\sigma(E_{th},E_\nu) = \int\frac{d\sigma(E_\nu,E_e)}{dE_e}dE_e
\label{CSmono}
\end{equation}
The limits obtained for various neutrino energies are summarized in Table~\ref{Table:Upper_Limits}. The obtained constraints are shown in Fig.~\ref{fig:Fig4} along with the results from  SuperKamiokande \cite{SK}. Borexino has set the best limits in the neutrino energy interval (0.5 -- 5)~MeV.

Since electron antineutrinos with energies greater than 1.8~MeV can interact with protons via IBD, we calculate their fluence upper limits for monoenergetic antineutrinos using relation (\ref{lim}) but replacing $N_e$ with number of protons $N_p$. 
The analysis is similar to a geo-neutrino search by Borexino based on $(5.5 \pm 0.3) \times 10^{31}$ protons $\times$ yr exposure.
Only $77$ antineutrino candidates were registered within $1842$ live-time days of data taking \cite{GEO-paper}. IBD interactions were detected by coincidence of a positron and then a delayed neutron with detection efficiency of $84.2 \pm 1.5$ \%. 
No IBD interactions were observed in $\pm 500$~s time windows around GW150914, GW151226 and GW170104 where the expected background is close to zero, so the 90\% C.L. upper limits on the number of  GW correlated events $N_{90}(E_\nu,n_{obs},n_{bkg})$ is $2.44$ \cite{Fel89}.
The IBD cross-section for antineutrinos was calculated according to \cite{Vissani}.
The results are shown on Fig.~\ref{fig:Fig4}, line 5 and in table~\ref{Table:Upper_Limits}, column~6.

\begin{figure}[ht!]
	\begin{center}
		\includegraphics[bb=80 100 450 750,width=0.8\linewidth]{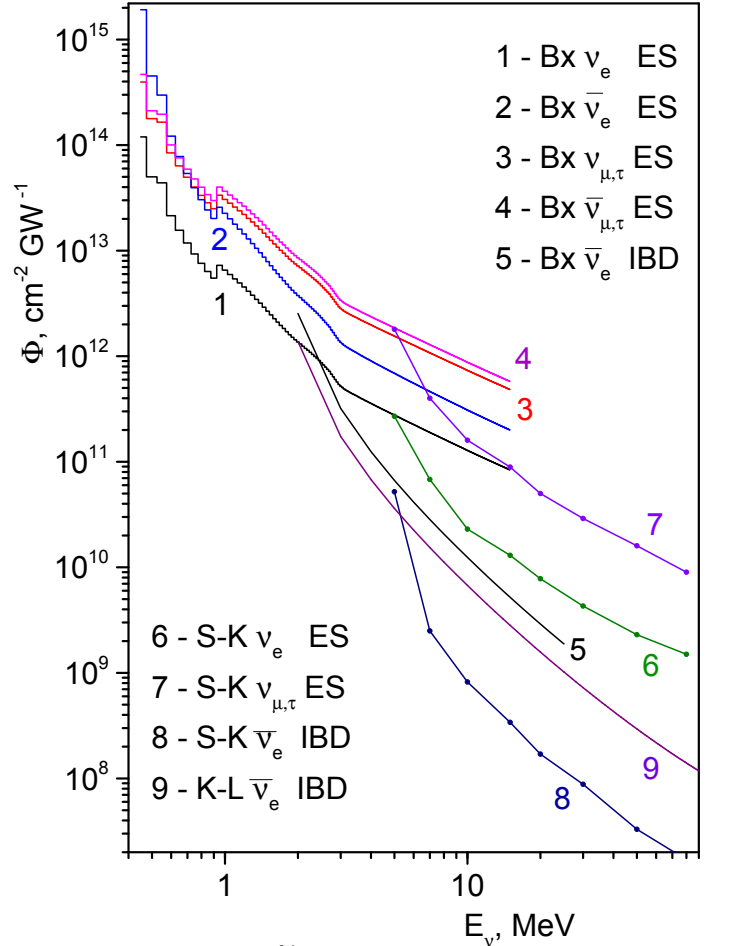}
		\caption{ Borexino 90\% C.L. fluence upper limits obtained through neutrino-electron elastic scattering for $\nu_e$ (line 1), $\bar{\nu_e}$ (line 2), $\nu_{\mu,\tau}$ (line 3), $\bar{\nu}_{\mu,\tau}$ (line 4) and through inverse beta-decay for $\bar{\nu_e}$ (line 5). Given are also the limits obtained by SuperKamiokande (line 6, 7, 8) and KamLAND (line 9).  }
		\label{fig:Fig4}
	\end{center}
\end{figure}

%Table 1. Upper limits on fluence of all neutrino flavours obtained through the study of neutrino-electron elastic scattering (90\% C.L).

\begin{table}[!t] 
	\begin{center}
		\begin{tabular}{|l|c|c|c|c|c|}
			\hline
			$E_\nu$, MeV& $\nu_e$ & $\nu_x$ & $\bar{\nu_e}$ & $\bar{\nu_x}$ & $\bar{\nu_e}$\:IBD   \\ \hline
			0.5    & 50      &  178    & 452             & 211             & -                      \\ \hline
			1.0    & 6.5     & 31      & 23              & 37              & -                      \\ \hline
			2.0    & 1.4     & 7.2      & 3.8             & 8.6              & 2.54                    \\ \hline
			3.0    & 0.52     & 2.8     & 1.4             & 3.4             & 0.32                   \\ \hline
			4.0    & 0.36     & 2.0     & 0.9             & 2.4             & 0.13                   \\ \hline
			5.0    & 0.28    & 1.6     & 0.69             & 1.9             & 0.067                   \\ \hline 
		\end{tabular}
		\vskip 12pt
		\caption{The upper limits on fluence per single GW event for all neutrino flavors in $10^{12}\:\rm{cm^{-2}}$ units at 90\% C.L. calculated for monochromatic neutrino lines}
		\label{Table:Upper_Limits}
	\end{center}
\end{table}

If the neutrino spectrum $\phi(E_\nu$) is not a monochromatic line, the total cross section for the electron recoil energy interval $(E_1,E_2)$ required for (\ref{lim}) is calculated as:

\begin{equation}
\sigma(E_1,E_2) = \int\int\frac{d\sigma(E_\nu,E_e)}{dE_e}\phi(E_\nu)dE_edE_\nu
\label{CSspectr}
\end{equation}

Since there is no reliable theory for the low-energy part of neutrino emission spectrum for BH-BH mergers, we calculate the fluence limits for two variants of neutrino spectrum $\phi(E_\nu)$. 
The first variant we considered is a standard power source model.
Since the neutrino energies that Borexino is sensitive to are relatively low, we drop the $E^{-2}$ dependence that is expected for high ($>$100 MeV) energy neutrinos and adopt the flat spectrum also used in \cite{SK}.
Additionally, we calculate the limits for the spectrum given by the normalized Fermi-Dirac (F-D) distribution for effective neutrino temperature $T$, connected with average neutrino energy as $\langle{E}\rangle \simeq 3$~$T$ and zero chemical potential ($\eta = 0$). 
\begin{equation}
\phi(E_\nu,T) \propto \frac{(E_\nu)^2}{1+\exp(E_\nu/T-\eta)}
\end{equation}

Although usage of the Fermi-Dirac distribution for approximation of the neutrino spectrum is only well motivated for a thermal neutrino flux (e.g. in SN collapse case \cite{Jan89,Kei03,Tam12}, whereas outflowing energy released during BH-BH mergers produces non-thermal radiation, it could still have a similar neutrino component.

Substituting the flat normalized distribution for neutrino energies between $0$ and $75$~MeV ($\phi(E_\nu) = const$) into (\ref{CSspectr}) and integrating over the analyzed electron recoil energy interval $(E_1,E_2) = (0.4, 15.0)$~MeV one gets the limits on the total electron neutrino fluence per single GW event:
\begin{equation}
\Phi(\nu_e) \leq 2.3\times 10^{10} \: \rm{cm^{-2}} 
\label{Flatlim}
\end{equation}
Limits obtained for other neutrino flavors are shown in Table~\ref{Table:Upper_Limits2}.

Limits on the fluence in the case of Fermi-Dirac distributions within the energy range $(0 - 500)$~MeV were calculated for different temperatures in steps of $0.5$~MeV.
The obtained limits are shown in Fig.~\ref{fig:Fig5}.
The obtained fluence constraints for the flat neutrino spectrum and Fermi-Dirac distribution with a temperature of $5$~MeV are shown in Table~\ref{Table:Upper_Limits2}. For comparison the limit on $\nu_e$ fluence in the case of a flat neutrino energy spectrum in the range ($3.5$ -- $75$)~MeV is $1.2\times10^9~\rm{cm^{-2}}$ \cite{SK} and the limit on  $\bar\nu_e$ fluence for F-D neutrino spectra at  $T$= 4 MeV is $3.6\times10^9 \rm{cm^{-2}}$ \cite{KL}.

\begin{table}[h] 
	\begin{center}
		\begin{tabular}{|l|c|c|c|c|c|}
			\hline
			Spectrum& $\nu_e$ & $\nu_x$ & $\bar{\nu_e}$ & $\bar{\nu_x}$ & $\bar{\nu_e}$\:IBD   \\ \hline
			Flat distribution    & 0.23      &  1.2    & 0.34             & 1.3             & 0.15                      \\ \hline
			F-D (T=5 MeV)       & 1.4     & 7.8      & 2.9              & 9.1              & 0.04                      \\ \hline
		\end{tabular}
		
		\vskip 12pt 
		\caption{The upper limits on GW event neutrino fluence in $10^{11}\:\rm{cm^{-2}}$ units for two cases of neutrino spectrum (90\% C.L.). Row 2 -- flat neutrino spectrum in the range $0 - 75$~MeV, Row 3 -- Fermi-Dirac spectrum for $T = 5$~MeV}
		\label{Table:Upper_Limits2}
	\end{center}
\end{table}

%These limits can be compared with S-K and KamLAND resukts

\begin{figure}[ht!]
	\begin{center}
		\includegraphics[bb= 90 80 489 760,width=0.8\linewidth]{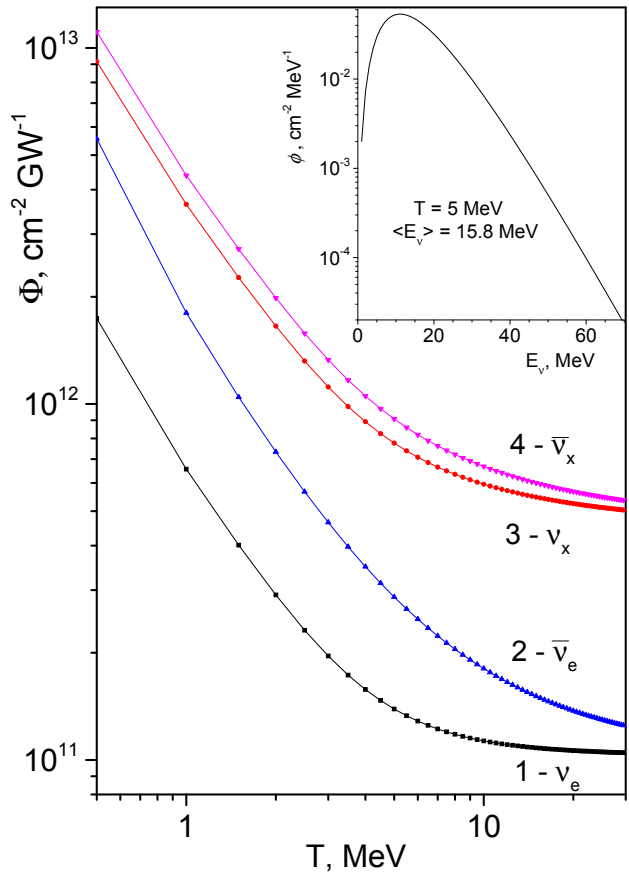}
		\caption{90\% C.L. upper limits on fluence for all neutrino flavors obtained for Fermi-Dirac neutrino spectrum with respect to effective neutrino temperature $T$. The inset shows Fermi-Dirac neutrino spectrum in case of $T = 5$~MeV. }
		\label{fig:Fig5}
	\end{center}
\end{figure}

The fluence upper limits can be converted into upper limits on the total energy radiated in the form of neutrinos for a BH-BH merger ($E\rm({BH\text{-}BH}\rightarrow\nu_{e,x},\bar{\nu}_{e,x})$).

Here, we consider only the energy radiated by electron neutrinos under the assumption of flat neutrino spectrum in the range (0-75) MeV and isotropic angular distribution. 
Usage of the LIGO-determined distance for GW150914, GW151226 and GW170104 and relation (\ref{Flatlim}) gives $E\rm{(BH\text{-}BH}\rightarrow\nu_e) \leq 4.0\times10^{61}$ erg.
This value could be compared with the energy emitted in the GW channel that is claimed to be around 2 solar masses per single GW, $2M_{\astrosun} $ = $3.6\times 10^{54}$ erg.
This suggests that successful detection of low-energy neutrinos should be possible only in the case of anisotropic angular distribution of neutrino emission. 
Limits on the energy radiated into neutrinos of other flavors can be easily calculated from table~\ref{Table:Upper_Limits2}.
\section{Conclusion}
We searched for an excess in the number of events detected by Borexino due to neutrino-electron scattering or IBD on protons correlated to the GW signals observed by the twin Advanced LIGO. We found no statistically significant increase in the number of events with an energy greater than 0.25 MeV in the detector during time windows of $\pm 500$~s around the  GW150914, GW151226 and GW170104 gravitational events. As a result, new limits on the fluence of monochromatic neutrinos of all flavors were set for neutrino energies (0.5--15) MeV. These are the strongest limits for $\nu_{e,\mu,\tau}$ and $\bar{\nu}_{\mu,\tau}$ for the neutrino energy range $(0.5 - 5.0)$~MeV and the constraint on electron antineutrino fluence based on $(\bar{\nu}_e,e)$-scattering is the strongest in the $(0.5 - 2.0)$~MeV energy range.
\section{Acknowledgments}
The Borexino program is made possible by funding from INFN (Italy); the NSF (U.S.); BMBF, DFG, (HGF, and MPI (Germany); RFBR (Grants No. 15-02-02117, No. 16-29-13014, No. 16-02-01026 and  No. 17-02-00305), RSF (Grant No. 17-12-01009)  (Russia); NCN Poland (Grant No. UMO-2013/10/E/ST2/00180); FNP Poland (Grant No. TEAM/2016-2/17). We acknowledge the generous support and hospitality of the Laboratori Nazionali del Gran Sasso (LNGS).
\end{section}

\end{document}